\documentclass[11pt,english]{article}
\usepackage[T1]{fontenc}
\usepackage[latin9]{inputenc}
\usepackage{geometry}
\usepackage{color}
\usepackage{babel}
\usepackage{amsmath}
\usepackage{amsthm}
\usepackage{setspace}
\usepackage[unicode=true,pdfusetitle,
 bookmarks=true,bookmarksnumbered=false,bookmarksopen=false,
 breaklinks=false,pdfborder={0 0 0},pdfborderstyle={},backref=false,colorlinks=false]
 {hyperref}

\makeatletter
\theoremstyle{definition}
\ifx\thechapter\undefined
  \newtheorem{example}{\protect\examplename}
\else
  \newtheorem{example}{\protect\examplename}[chapter]
\fi
\theoremstyle{plain}
\ifx\thechapter\undefined
  \newtheorem{prop}{\protect\propositionname}
\else
  \newtheorem{prop}{\protect\propositionname}[chapter]
\fi
\theoremstyle{plain}
\ifx\thechapter\undefined
  \newtheorem{thm}{\protect\theoremname}
\else
  \newtheorem{thm}{\protect\theoremname}[chapter]
\fi
\theoremstyle{definition}
\ifx\thechapter\undefined
  \newtheorem{defn}{\protect\definitionname}
\else
  \newtheorem{defn}{\protect\definitionname}[chapter]
\fi

\usepackage{colortbl}
\usepackage{tikz}
\usetikzlibrary{backgrounds,calc,positioning}
\date{}

\makeatother

\providecommand{\definitionname}{Definition}
\providecommand{\examplename}{Example}
\providecommand{\propositionname}{Proposition}
\providecommand{\theoremname}{Theorem}

\begin{document}
\title{Slot-specific Priorities with Capacity Transfers\thanks{First version: April, 2018. This version: September, 2020.\protect \\
The current version of the paper subsumes and supersedes Avataneo's
undergraduate thesis that was submitted to ITAM in 2018. We would
like to thank Orhan Aygün, Jenna M. Blochowicz, Diego Domínguez, Federico
Echenique, Andrei Gomberg, Ravi Jagadeesan, Onur Kesten, Scott Duke
Kominers, Sera Linardi, Romans Pancs, Ali O\u{g}uz Polat, and Rakesh
Vohra, as well as the seminar participants at ITAM and Carnegie Mellon
University for their helpful comments and suggestions. }}
\author{Michelle Avataneo\thanks{Avataneo: Kellogg School of Management, Northwestern University, 2211
Campus Dr, Evanston, IL, 60208, USA (Email: mavataneotruqui@gmail.com).} $\quad$and $\quad$Bertan Turhan\thanks{Turhan: Department of Economics, Iowa State University, Heady Hall,
518 Farm House Ln, Ames, IA, 50011, USA (Email: bertan@iastate.edu).} }
\date{\bigskip{}
}
\maketitle
\begin{abstract}
In many real-world matching applications, there are restrictions for
institutions either on priorities of their slots or on the transferability
of unfilled slots over others (or both). Motivated by the need in
such real-life matching problems, this paper formulates a family of
practical choice rules, \emph{slot-specific priorities with capacity
transfers} (SSPwCT). These practical rules invoke both slot-specific
priorities structure and transferability of vacant slots. We show
that the \emph{cumulative offer mechanism} (COM) is\emph{ stable},\emph{
strategy-proof} and \emph{respects improvements} with regards to SSPwCT
choice rules. Transferring the capacity of one more unfilled slot,
while all else is constant, leads to \emph{strategy-proof} \emph{Pareto
improvement} of the COM. Following \emph{Kominer'}s\emph{ }(2020)
formulation, we also provide comparative static results for expansion
of branch capacity and addition of new contracts in the SSPwCT framework.
Our results have implications for resource allocation problems with
diversity considerations. 

\bigskip{}

\emph{$\mathbf{JEL}$ $\mathbf{classification}$: }C78, D47

\emph{$\mathbf{Keywords}$: }Market design, matching, affirmative
action 

\vfill{}
\end{abstract}
\pagebreak{}

\section{Introduction}

The slot-specific priorities framework of\emph{ Kominers and Sönmez}
(2016) is an influential many-to-one matching model with contracts.
Each contract is between an agent and an institution, and specifies
some terms and conditions. Slots have their own (potentially independent)
rankings for contracts. Within each institution,  a linear order --
referred to as the \emph{precedence order -- }determines the sequence
in which slots are filled. This framework provides a powerful tool
for market designers and policymakers to handle diversity and affirmative
action. Institutions choose contracts by filling their slots sequentially.
An agent might have different priority at different slots of the same
institution . In the context of cadet-branch matching (\emph{Switzer
and Sönmez}, 2013 and \emph{Sönmez}, 2013), for example, some slots
for each service branch grant higher priority for cadets who are willing
to serve additional years of service.\emph{ Kominers and Sönmez} (2016)
develop a general framework to handle these types of slot-specific
priority structures. The slot-specific priorities framework embeds
the settings of the following works among others: \emph{Balinski and
Sönmez} (1999), \emph{Abdulkadiro\u{g}lu and Sönmez} (2003), \emph{Kojima}
(2012), \emph{Hafalir et al.} (2013), and \emph{Aygün and Bó} (2020),
\emph{Switzer and Sönmez} (2013), and \emph{Sönmez} (2013). 

The slot-specific priorities offer flexible solutions to important
real-world matching problems. \emph{Aygün and Bó }(2020), for example,
design slot-specific priorities choice rules for the Brazilian college
admission problem, where students have multidimensional privileges.
In 2012 Brazilian public universities were mandated to use affirmative
action policies for candidates from racial and income minorities.
The law established that certain fractions of the accepted students
in each program should have studied in public high schools, come from
a low-income family, and/or belong to racial minorities. This objective
was implemented by partitioning the positions in each program, earmarking
them for different combinations of these affirmative action characteristics.
\emph{Aygün and Bó} (2020) analyze several of these choice rules,
present their shortcomings and correct their shortcomings by designing
slot-specific priorities choice rules for programs.

More recently, \emph{Pathak et al. }(2020a) invoke the slot-specific
priorities framework to design triage protocol for ventilator rationing.\footnote{The most recent version of \emph{Pathak et al. }(2020a) considers
more general framwork. See \emph{Pathak et al. }(2020b). } The authors analyze the consequences of using a reserve system---in
which ventilators are partitioned into multiple categories---for
rationing medical resources. The authors propose sequential reserve
matching rules, which are first introduced in the slot-specific framework
of\emph{ Kominers and Sönmez} (2016). 

When slot priorities are restricted due to institutional constraints,
the full potential of the slot-specific priorities framework might
not be achieved. Another useful tool that allows interactions across
slots is \emph{capacity transfers} ---introduced by \emph{Westkamp}
(2013). Capacity transfers allow more general forms of interactions
across slots than slot specific priorities do. However, \emph{Westkamp}
(2013) does not allow the variation in contractual terms. For applications
such as cadet-branch matching (\emph{Switzer and Sönmez}, 2013 and
\emph{Sönmez}, 2013), airline upgrade allocation (\emph{Kominers and
Sönmez}, 2016), and admissions for publicly funded educational institutions
and government-sponsored jobs in India (\emph{Aygün and Turhan}, 2017,
2020a, 2020b), allowing the variation in contractual terms is necessary. 

In many real-world applications, either slot priorities (for some
slots if not all) or the ability to transfer vacant slots may be restricted
due to institutional constraints especially in diversity and affirmative
action in school choice, college admissions, government-sponsored
job matching, and also faculty hiring. When both are restricted, some
slots might remain unfilled. This, in turn, might lead to a Pareto
inferior outcome.  A real-world example of this case can be seen in
India. For admissions to publicly funded educational institutions
and government-sponsored jobs in India, each institution reserves
15 percent of its slots for people from Scheduled Castes (SC), 7.5
percent for people from Scheduled Tribes (ST), and 27 percent for
people from Other Backward Classes (OBC). People who do not belong
to any of these categories are referred to as members of the \emph{General
Category }(GC). The remaining slots are called\emph{ open-category}
slots and are available to everyone, including those from SC, ST,
and OBC. In each institution, for slots that are reserved for SC,
ST and OBC, only applicants who declare they belong to these respective
categories are considered. If there is low demand from either SC or
ST applicants, some slots remain vacant. Vacant SC/ST slots can potentially
be utilized by two ways: (1) other candidates can be made eligible
for SC/ST slots, but at a lower priority than all SC/ST applicants,
or (2) vacant SC/ST slots can be reverted into, say, open-category
slots. Currently, none of these possibilities are allowed. Each year
many SC/ST slots remain vacant. 

There are instances where slot priorities are restricted but transferability
is allowed.\emph{ Baswana et al.} (2018) designed and implemented
a new joint seat allocation process for technical universities in
India. Since 2008, following a Supreme Court decision, unfilled OBC
slots are required to be made available to GC candidates of publicly
funded educational institutions. However, institutions are \emph{prohibited}
from modifying the priorities of OBC slots. This possibility was offered
to Indian authorities but ultimately rejected. On the other hand,
reverting vacant OBC slots into GC slots is allowed.\emph{ Baswana
et al}. (2018) report their interaction with the Indian policy makers
as follows: 
\begin{quote}
``Business rule 5 required unfilled OBC seats to be made available
to Open category candidates. \emph{The approach we initially suggested
involved construction of augmented Merit Lists making Open category
candidates eligible for OBC seats but at a lower priority than all
OBC candidates}, and modification of virtual preference lists so that
general candidates now apply for both the OPEN and the OBC virtual
programs. We showed that running our algorithm on these modified inputs
would produce the candidate optimal allocation satisfying the business
rules. However, \textcolor{black}{\emph{the authorities feared that
this approach may cause issues}} with computing the closing rank correctly
(see Design Insight 6), or have some other hidden problem. An authority
running centralized college or school admissions is typically loathe
to modify, add complexity to, or replace software that is tried and
tested.'' 
\end{quote}
In this paper, we invoke both of these powerful tools --i.e., capacity
transfers across slots and independent slot priorities-- to formulate
a larger set of practical choice rules. Our aim is to expand the toolkit
of market designers to be able to implement comprehensive selection
criteria, especially when there are institutional restrictions. We
construct institutional choices as follows: Each institution has two
types of slots: \emph{original} slots and \emph{shadow} slots. Each
shadow slot has an initial capacity of 0 and is associated with an
original slot. Each slot (original and shadow) has a linear order
(potentially independent) over contracts. Institutions have precedence
orders over original and shadow slots. Each institution  is also endowed
with a \emph{location vector} for shadow slots. The exact order at
which slots are processed is determined by the precedence orders over
original and shadow slots together with the location vector. The interaction
between associated original and shadow slots is as follows: If an
original slot is assigned a contract, then the capacity of the associated
shadow slot remains 0, i.e., the shadow slot will be inactive. If
an original slot cannot be filled, the institution  has the option
to transfer its capacity to the associated shadow slot. In this case,
the capacity of the shadow slot becomes 1. A capacity transfer vector
of the institution  determines for which original slots such reversion
is allowed and for which ones it is not. 

When the transferability of all original slots is prohibited, our
model reduces to that of \emph{Kominers and Sönmez }(2016). Given
the exact precedence order (i.e., precedence orders over original
and shadow slots, location vector of shadow slots), capacity transfer
vector, and slot priorities over contracts, slots are filled \emph{sequentially}
in a straightforward manner. We call this family of choice rules that
are constructed this way \emph{Slot-specific Priorities with Capacity
Transfers Choice Rules }(SSPwCT). We show how markets with SSPwCT
choice rules can be cleared by the COM. To do so, we borrowed the
hidden substitutes theory of \emph{Hatfield and Kominers (2019)}.
The COM is the unique stable and strategy-proof mechanism in the SSPwCT
environment (Theorem 1) that also respects improvements (Theorem 2).

Finally, we provide several comparative static regarding the outcome
of the COM  with respect to SSPwCT choice rules. We first show that
when an institution  reverts one more original slot into a shadow
slot in the case of a vacancy, if all else is fixed, we obtain a strategy-proof
Pareto improvement under the COM (Theorem 3). Then, building on \emph{Kominers'
}(2020) analysis, we show that the outcome of the COM  is (weakly)
improved for all agents when (1) an original slot is added to an institution,
 while all else remains fixed (Theorem 4), (2) new contracts are added
at the bottom of slots' priority orders right before the null contract
(Theorem 5), and (3) the new contracts of a single agent are added
anywhere in the slots' priority orders (Theorem 6).

SSPwCT choice rules may be used in cadet-branch matching in USMA and
ROTC (\emph{Sönmez and Switzer, }2013 and \emph{Sönmez, }2013), resource
allocation in India with comprehensive affirmative action (\emph{Boswana
et al.,} 2019, \emph{Aygün and Turhan, }2020a\emph{ }and\emph{ }2020b),
Chilean school choice and college admissions with affirmative action
(\emph{Rios et al., }2018 and \emph{Correa et al., 2019})\emph{ },
and Brazilian college admissions with multidimensional reserves (\emph{Aygün
and Bó}, 2020). 

\subsection*{Related Literature}

Our model is built on that of \emph{Kominers and Sönmez} (2016). The
SSPwCT choice rules expands the slot-specific priorities choice rules
of \emph{Kominers and Sönmez} (2016) by allowing transferability of
vacant slots. When transferability is not allowed, the SSPwCT choice
rules are equivalent to the slot-specific priorities choice rules.
As opposed to \emph{Kominers and Sönmez} (2016), in the presence of
capacity transfers, it is not possible to define associated one-to-one
model of agent-slot market in our setting. We instead invoke \emph{Hatfield
and Kominers' }(2019) \emph{hidden substitutes} theory to show that
the COM with respect to SSPwCT choice rules is stable and strategy-proof. 

Recently, \emph{Aygün and Turhan} (2020) show that each slot-specific
priorities choice rule can be written as a \emph{dynamic reserves
choice rule}. The family of SSPwCT choice rules embeds \emph{Aygün
and Turhan}'s (2020) dynamic reserves choice rules family. There are
two important advantages of the SSPwCT rules over the dynamic reserves
choice rules. The first one is that the SSPwCT functions are simpler
to describe. The capacity functions in the setting of dynamic reserves
are complicated, whereas transfer functions in the SSPwCT environment
is just a vector of binary variables. The second one is that in the
dynamic reserves setting there is a master priority ordering for each
institution, and the priority orderings of different groups of slots
within the institution are obtained from the master priority ordering
in a straightforward manner by removing some agents. In the current
paper, following the model of \emph{Kominers and Sönmez} (2016), slots
have more general priorities (possibly independent) over contracts. 

\emph{Kominers (2020)} gives a novel proof of the entry comparative
static via the respecting improvement property, and also shows that
his proof extends to yield comparative static results in matching
with slot-specific priorities framework. By adapting his notation
and formulation, we further extend\emph{ }his results to the SSPwCT
environment. Papers that study entry comparative static include \emph{Kelso
and Crawford }(1982), \emph{Gale and Sotomayor }(1985), \emph{Crawford
}(1991), \emph{Hatfield and Milgrom }(2005), \emph{Biró et al.} (2008),
\emph{Ostrovsky} (2008),  and \emph{Chambers and Yenmez} (2017), among
others. 

Our paper uses the matching with contracts framework that is introduced
by \emph{Fleiner} (2003) and \emph{Hatfield and Milgrom} (2005). Important
work on matching with contracts, among many others, include \emph{Hatfield
and Kojima} (2010), \emph{Aygün and Sönmez} (2013), \emph{Afacan}
(2017), \emph{Hatfield and Kominers (2019)}, and \emph{Hatfield et
al.} (2017, 2019).

We study a family of lexicographic choice rules with transfers. Lexicographic
choice rules (without transfers) are recently studied in\emph{ Chambers
and Yenmez }(2018) and \emph{Do\u{g}an et al. }(2018), among others.\emph{ }

Finally, our paper also contributes the matching literature on distributional
constraints. There is extensive literature on this line of literature.
A partial list includes \emph{Abdulkadiro\u{g}lu and Sönmez} (2003),
\emph{Abdulkadiro\u{g}lu }(2005), \emph{Biró et al.} (2010), \emph{Kojima}
(2012), \emph{Hafal\i r et al.} (2013), \emph{Westkamp }(2013)\emph{,
Echenique and Yenmez} (2015)\emph{, Kamada and Kojima} (2015, 2017,
and 2018), \emph{Fragiadakis and Troyan} (2017), \emph{Aygün and Turhan
(2017, 2020a, 2020b), Jagadeesan }(2019), \emph{Nguyen and Vohra }(2019),
\emph{Afacan }(2020), and \emph{Dur et al. }(2020). 

\section{Matching with Contracts Framework }

There is a finite set of \emph{agents} $\mathcal{I}=\{i_{1},...,i_{n}\}$
and a finite set of\emph{ branches} $\mathcal{B}=\{b_{1},...,b_{m}\}$.
There is a finite set of \emph{contracts }$\mathcal{X}$. Each contract
$x\in\mathcal{X}$ is associated with an agent $\mathbf{i}(x)$ and
a branch $\mathbf{b}(x)$. There may be many contracts for each agent-branch
pair. We call a set of contracts $X\subseteq\mathcal{X}$ an \emph{outcome},
with $\mathbf{i}(X)\equiv\underset{x\in X}{\cup}\{\mathbf{i}(x)\}$
and $\mathbf{b}(X)\equiv\underset{x\in X}{\cup}\{\mathbf{b}(x)\}$.
For any $i\in\mathcal{I}$, we let $X_{i}\equiv\{x\in X\mid\mathbf{i}(x)=i\}$.
Similarly, for any $b\in\mathcal{B}$, we let $X_{b}\equiv\{x\in X\mid\mathbf{b}(x)=b\}$.

Each agent $i\in\mathcal{I}$ has unit demand over contracts in $\mathcal{X}_{i}\equiv\{x\in\mathcal{X}\mid\mathbf{i}(x)=i\}$
and an outside option $\emptyset_{i}$. The strict preference of agent
$i$ over $\mathcal{X}_{i}\cup\{\emptyset_{i}\}$ is denoted by $P_{i}$.
A contract $x\in\mathcal{X}_{i}$ is \emph{acceptable for $i$ }(with
respect to $P_{i}$) if $xP_{i}\emptyset_{i}$. Agent preferences
over contracts are naturally extended to preferences over outcomes.
For each individual $i\in\mathcal{I}$ and set of contracts $X\subseteq\mathcal{X}$,
we denote by $\underset{P_{i}}{max}X$ the maximal element of $X$
according to preference order $P_{i}$, and we assume that $\underset{P_{i}}{max}X=\emptyset$
if $\emptyset_{i}P_{i}x$ for all $x\in X$. Each individual always
chooses the best available contract according to his preferences,
so that choice rule $C^{i}(X)$ of an individual $i\in\mathcal{I}$
from contract set $X\subseteq\mathcal{X}$ is defined by $C^{i}(X)=\underset{P_{i}}{max}X$. 

Each branch $b\in\mathcal{B}$, on the other hand, has multi-unit
demand and is endowed with a choice rule $C^{b}$. We let $q_{b}$
denote the capacity of branch $b$. The choice rule $C^{b}$ describes
how branch $b$ would choose from any offered set of contracts. We
assume throughout that for all $X\subseteq\mathcal{X}$ and for all
$b\in\mathcal{B}$, the choice rule $C^{b}$ satisfies $(1)$ $C^{b}(X)\subseteq X_{b}$,
and $(2)$ $C^{b}(X)$ is feasible in the sense that it contains at
most one contract of any given agent. For any $X\subseteq\mathcal{X}$
and $b\in\mathcal{B}$, we denote by $R^{b}(X)\equiv X\setminus C^{b}(X)$
the set of contracts that $b$ \emph{rejects }from $X$. 

A set of contracts $X\subseteq\mathcal{X}$ is a \emph{feasible} \emph{outcome}
if $\mid X_{i}\mid\leq1$, for every $i\in\mathcal{I}$, and $\mid X_{b}\mid\leq q_{b}$,
for each $b\in\mathcal{B}$. 

A feasible outcome $X\subseteq\mathcal{X}$ is $\mathbf{stable}$
if 
\begin{enumerate}
\item $C^{i}(X)=X_{i}$ for all $i\in\mathcal{I}$, and $C^{b}(X)=X_{b}$,
for all $b\in\mathcal{B}$, and 
\item there is no $b\in\mathcal{B}$ and set of contracts $Y\neq C^{b}(X)$
such that $Y_{b}=C^{b}(X\cup Y)$ and $Y_{i}=C^{i}(X\cup Y)$ for
all $i\in\mathbf{i}(Y)$. 
\end{enumerate}
The first condition is known as \emph{individual rationality }and
requires that no agent or branch would rather drop one of the signed
contracts. The second condition is the \emph{no blocking }requirement.
If fails, then there is an alternative set of contracts that a branch
and agents associated with a contract in that set strictly prefers. 

\subsection{Slot-specific Priorities with Capacity Transfers (SSPwCT) Choice
Rules}

Each branch $b\in\mathcal{B}$ has two types of seats: $\mathbf{original}$
seats and $\mathbf{shadow}$ seats. Let $O_{b}=\{o_{b}^{1},o_{b}^{2},...,o_{b}^{n_{b}}\}$
and $E_{b}=\{e_{b}^{1},e_{b}^{2},...,e_{b}^{n_{b}}\}$ be branch $b$'s
set of original seats and shadow seats, respectively, where $n_{b}$
denotes the physical capacity of branch $b$. Each seat in both $O_{b}$
and $E_{b}$ has priority orders over contracts in $\mathcal{X}_{b}\cup\{\emptyset\}$
denoted by $\Pi_{b}^{o}$ for $o\in O_{s}$ and $\Pi_{b}^{e}$ for
$e\in E_{s}$ (the weak orders are denoted by $\Gamma^{o}$ and $\Gamma^{e}$)
and can be assigned at most one contract in $\mathcal{X}_{b}\equiv\{x\in\mathcal{X}\mid\mathbf{b}(x)=b\}$.
Let $\Pi_{b}=(\Pi_{b}^{o},\Pi_{b}^{e})$ denote the priority profile
of branch $b$ and $\Pi=(\Pi_{b}){}_{b\in\mathcal{B}}$ denote the
priority profiles of all branches. We denote by $\underset{\pi^{o}}{max}X$
the maximal element of $X$ according to priority ordering $\Pi^{o}$
and by $\underset{\pi^{e}}{max}X$ the maximal element of $X$ according
to priority ordering $\Pi^{e}$. We assume $\underset{\pi^{o}}{max}X=\emptyset$
if $\emptyset\Pi^{o}x$ for all $x\in X$ and $\underset{\pi^{e}}{max}X=\emptyset$
if $\emptyset\Pi^{e}x$ for all $x\in X$. 

Each branch $b\in\mathcal{B}$ has two linear precedence orders, one
over original seats, $\triangleright_{b}^{O}$ , and one over shadow
seats, $\triangleright_{b}^{E}$. We denote $O_{b}=\{o_{b}^{1},o_{b}^{2}...,o_{b}^{n_{b}}\}$
with $o_{b}^{l}\triangleright_{b}^{O}o_{b}^{l+1}$ and $E_{b}=\{e_{b}^{1},e_{b}^{2}...,e_{b}^{n_{b}}\}$
with $e_{b}^{l}\triangleright_{b}^{E}e_{b}^{l+1}$ unless otherwise
stated. The interpretation of $\triangleright_{b}^{O}$ is that if
$o\triangleright_{b}^{O}o'$, then, whenever possible, branch $b$
fills seat $o$ before $o^{'}$. Each shadow seat is associated with
an original seat. If the original seat remains empty, then branch
$b$ can decide whether to transfer its capacity to its associated
shadow seat, which initially has no capacity, through a capacity transfer
scheme $q_{b}$ defined below.

A \emph{capacity transfer scheme} is an integer-valued vector $q_{b}=(q_{b}^{1},q_{b}^{2},...,q_{b}^{n_{b}})$
such that for every $k=1,...,n_{b}$:
\[
q_{b}^{k}=\begin{cases}
0 & \text{ \text{if branch \ensuremath{b} does not transfer capacity from \ensuremath{o_{b}^{k}} to \ensuremath{e_{b}^{k}} when \ensuremath{o_{b}^{k}} is not filled. } }\\
1 & \text{ if branch \ensuremath{b} transfers capacity from \ensuremath{o_{b}^{k}}to \ensuremath{e_{b}^{k}} when \ensuremath{o_{b}^{k}} is not filled. }
\end{cases}
\]

Since a capacity transfer from $o_{b}^{k}$ to $e_{b}^{k}$ is possible
\emph{only when} $o_{b}^{k}$ is not filled, the physical capacity
of branch $b$ is never violated. We define an \emph{indicator function}
for the original seats as follows: 
\[
\mathbf{1}_{o_{b}^{l}}=\begin{cases}
0 & \text{ \text{if seat \ensuremath{o_{b}^{l}} remains empty.} }\\
1 & \text{ if seat \ensuremath{o_{b}^{l}} is filled.}
\end{cases}
\]

Given precedence orders $\triangleright_{b}^{O}$ and $\triangleright_{\text{b}}^{E}$,
a \emph{location vector} for the shadow seats of branch $b$ is an
integer-valued vector $L_{b}=(l_{1},...,l_{n_{b}})$ where $l_{k}$
is the number of original seats that precede shadow seat $e_{b}^{k}$
that satisfy the following condition: 
\[
L_{b}=\{(l_{1},...,l_{n_{b}})\mid k\leq l_{k}\leq n_{b}\;\forall k=1,...,n_{b}\text{ and }l_{k}\geq l_{k-1}\;\forall k=2,...,n_{b}\}.
\]

The condition in the definition of $L_{b}$ ensures that for every
shadow seat, the number of preceding original seats is greater than
the number of preceding shadow seats. Hence, a shadow seat will never
come before its associated original seat in this order. The location
vector $L_{b}$ together with precedence orders $\triangleright_{b}^{O}$
and $\triangleright_{b}^{E}$ gives us the exact order in which the
original and shadow seats are filled. Let $\triangleright_{b}\equiv(\triangleright_{b}^{O},\triangleright_{b}^{E},L_{b})$
denote the exact order of branch $b$'s slots. We illustrate this
with an example below. 
\begin{example}
Consider a branch with three original seats with $o_{b}^{1}\triangleright_{b}^{O}o_{b}^{2}\triangleright_{b}^{O}o_{b}^{3}$
and three shadow seats with $e_{b}^{1}\triangleright_{b}^{E}e_{b}^{2}\triangleright_{b}^{E}e_{b}^{3}$
together with the location vector $L_{b}=(1,3,3)$. The order $\triangleright_{b}$
in which the original and shadow seats are filled is as follows: 
\end{example}
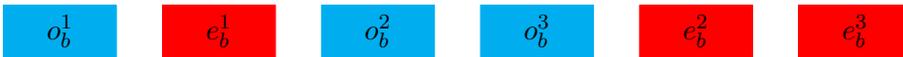
\begin{tikzpicture} [align=center, node distance =6mm]
\node[shape=rectangle, draw=cyan, fill=cyan, minimum width=1.5cm] (firstbox) {$o_{b}^{1}$};  
\node[shape=rectangle, draw=red, fill=red, minimum width=1.5cm] (secondbox) [right=of firstbox] {$e_{b}^{1}$};
\node[shape=rectangle, draw=cyan, fill=cyan, minimum width=1.5cm] (thirdbox) [right=of secondbox] {$o_{b}^{2}$};    
\node[shape=rectangle, draw=cyan, fill=cyan, minimum width=1.5cm] (fourthbox) [right=of thirdbox] {$o_{b}^{3}$}; 
\node[shape=rectangle, draw=red, fill=red, minimum width=1.5cm] (fifthbox) [right=of fourthbox] {$e_{b}^{2}$};  
\node[shape=rectangle, draw=red, fill=red, minimum width=1.5cm] (sixthbox) [right=of fifthbox] {$e_{b}^{3}$};  

\end{tikzpicture}

\subsection*{Description of SSPwCT Choice Rules}

For branch $b\in\mathcal{B}$, $C^{b}(\cdot,\triangleright_{b},q_{b},\Pi_{b}):2^{X}\rightarrow2^{X}$
denotes the choice rule of branch $b$ given the precedence order
of slots $\triangleright_{b}$, the capacity transfer function $q_{b}$,
and the priority profile of slots $\Pi_{b}$. Given a set of contracts
$X\subseteq\mathcal{X}$, $C^{b}(X,\triangleright_{b},q_{b},\Pi_{b})$
denotes the set of chosen contracts for branch $b$ from the set of
contracts $X$. 

To formulate the choice rule, we first rename the slots as $S=(s^{1},s^{2},...,s^{2n_{b}}$)
where $s^{k}$ is either an original or a shadow seat, depending on
$\triangleright_{b}^{O}$ , $\triangleright_{\text{b}}^{E}$ , and
$L_{b}$. In Example 1 above with $L_{b}=(1,3,3)$, we can rename
slots as follows: $s_{b}^{1}=o_{b}^{1}$, $s_{b}^{2}=e_{b}^{1},$$s_{b}^{3}=o_{b}^{2}$,
$s_{b}^{4}=o_{b}^{3},$ $s_{b}^{5}=e_{b}^{2}$, and $s_{b}^{6}=e_{b}^{3}$.
It is important to note that $\Pi_{b}^{s_{b}^{1}}=\Pi_{b}^{o_{b}^{1}}$,
$\Pi_{b}^{s_{b}^{2}}=\Pi_{b}^{e_{b}^{1}}$, etc... 

We are now ready to describe the choice procedure. Given $X\subseteq\mathcal{X}$: 
\begin{itemize}
\item Start with the original seat $s_{b}^{1}$. Assign the contract $x^{1}$
that is $\Pi_{b}^{s_{b}^{1}}-maximal$ among the contracts in $X$. 
\item If $s_{b}^{2}$ is either an original or a shadow seat such that $\mathbf{1}_{o_{b}^{1}}=0$
and $q_{b}^{1}=1$, assign the contract $x^{2}$ that is $\Pi_{b}^{s_{b}^{2}}-maximal$
among the contracts in $X\setminus X_{\mathbf{i}(\{x^{1}\})}$. Otherwise,
assign the empty set.
\item This process continues in sequence. If $s_{b}^{k}$ is an original
seat or a shadow seat such that $\mathbf{1}_{o_{b}^{r}}=0$, where
$o_{b}^{r}$ is the original seat that is associated with the shadow
seat $s_{b}^{k}$, and $q_{b}^{r}=1$, then assign contract $x^{k}$
that is $\Pi_{b}^{s_{b}^{k}}-maximal$ among the contracts in $X\setminus X_{\mathbf{i}(\{x^{1},...,x^{k-1}\})}$.
Otherwise, assign the empty set.
\end{itemize}
Given $n_{b}$, $(\triangleright_{b},q_{b},\Pi_{b})$ parametrizes
the family of SSPwCT choice rules for branch $b$. 

\subsubsection*{Examples of SSPwCT Choice Rules }
\begin{example}
Consider $b\in\mathcal{B}$ with $n_{b}=3$, $L_{b}=(1,2,3)$, and
$q_{b}=(1,1,1)$. The capacity transfer scheme allows branch $b$
to transfer capacities from original seats to shadow seats whenever
they remain unfilled. Given the location vector and capacity transfer
scheme, the choice procedure for branch $b$ is as follows. Given
an offer set $X\subseteq\mathcal{X}$: 
\end{example}
\begin{itemize}
\item Assign $o_{b}^{1}$ the contract $x^{1}$ that is $\Pi_{b}^{o_{b}^{1}}-maximal$
among the contracts in $X$. 
\item If $\mathbf{1}_{o_{b}^{1}}=0$, assign $e_{b}^{1}$ the contract $x^{2}$
that is $\Pi_{b}^{e_{b}^{1}}-maximal$ among the contracts in $X\setminus X_{\mathbf{i}(\{x^{1}\})}$.
Otherwise, assign $e_{b}^{1}$ the empty set.
\item Assign $o_{b}^{2}$ the contract $x^{3}$ that is $\Pi_{b}^{o_{b}^{2}}-maximal$
among the contracts in $X\setminus X_{\mathbf{i}(\{x^{1},x^{2}\})}$. 
\item If $\mathbf{1}_{o_{b}^{2}}=0$, assign $e_{b}^{2}$ the contract $x^{4}$
that is $\Pi_{b}^{e_{b}^{2}}-maximal$ among the contracts in $X\setminus X_{\mathbf{i}(\{x^{1},x^{2},x^{3}\})}$.
Otherwise, assign $e_{b}^{2}$ the empty set.
\item Assign $o_{b}^{3}$ the contract $x^{5}$ that is $\Pi_{b}^{o_{b}^{3}}-maximal$
among the contracts in $X\setminus X_{\mathbf{i}(\{x^{1},x^{2},x^{3},x^{4}\})}$. 
\item If $\mathbf{1}_{o_{b}^{3}}=0$, assign $e_{b}^{3}$ the contract $x^{6}$
that is $\Pi_{b}^{e_{b}^{3}}-maximal$ among the contracts in $X\setminus X_{\mathbf{i}(\{x^{1},x^{2},x^{3},x^{4},x^{5}\})}$.
Otherwise, assign $e_{b}^{3}$ the empty set.
\end{itemize}
The following picture depicts the order of slots and the capacity
transfer scheme where arrows indicate that the capacity is transferred
if the original seat remains empty:

\medskip{}
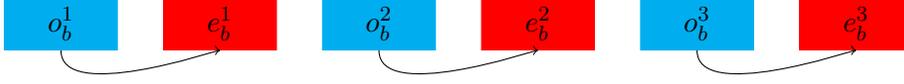
\begin{tikzpicture} [align=center, node distance =6mm]
\node[shape=rectangle, draw=cyan, fill=cyan, minimum width=1.5cm] (firstbox) {$o_{b}^{1}$};  
\node[shape=rectangle, draw=red, fill=red, minimum width=1.5cm] (secondbox) [right=of firstbox] {$e_{b}^{1}$};
\node[shape=rectangle, draw=cyan, fill=cyan, minimum width=1.5cm] (thirdbox) [right=of secondbox] {$o_{b}^{2}$};  
\node[shape=rectangle, draw=red, fill=red, minimum width=1.5cm] (fourthbox) [right=of thirdbox] {$e_{b}^{2}$};  
\node[shape=rectangle, draw=cyan, fill=cyan, minimum width=1.5cm] (fifthbox) [right=of fourthbox] {$o_{b}^{3}$}; 
\node[shape=rectangle, draw=red, fill=red, minimum width=1.5cm] (sixthbox) [right=of fifthbox] {$e_{b}^{3}$};  
\draw[->] (firstbox.south) .. controls +(down:7mm) and +(left:.5mm).. (secondbox.south);    
\draw[->] (thirdbox.south) .. controls +(down:7mm)and +(left:.5mm).. (fourthbox.south);  
\draw[->] (fifthbox.south) .. controls +(down:7mm) and +(left:.5mm)..(sixthbox.south);  
\end{tikzpicture}

In the previous example each shadow seat appears right after its corresponding
original seat. It is common in practice for the institution to try
to fill its original seats first before the shadow seats. We provide
such an example below. 
\begin{example}
Consider branch $b\in\mathcal{B}$ with $n_{b}=3$, $L_{b}=(3,3,3)$,
and $q_{b}=(1,1,1)$. The following picture depicts the order of slots
and the capacity transfer scheme where arrows indicate that the capacity
is transferred if the original seat remains empty:
\end{example}
\medskip{}

\medskip{}

\medskip{}

\medskip{}
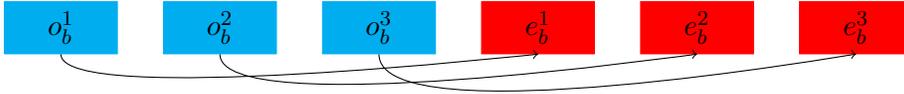
\begin{tikzpicture} [align=center, node distance =6mm]
\node[shape=rectangle, draw=cyan, fill=cyan, minimum width=1.5cm] (firstbox) {$o_{b}^{1}$};  
\node[shape=rectangle, draw=cyan, fill=cyan, minimum width=1.5cm] (secondbox) [right=of firstbox] {$o_{b}^{2}$};
\node[shape=rectangle, draw=cyan, fill=cyan, minimum width=1.5cm] (thirdbox) [right=of secondbox] {$o_{b}^{3}$};    
\node[shape=rectangle, draw=red, fill=red, minimum width=1.5cm] (fourthbox) [right=of thirdbox] {$e_{b}^{1}$}; 
\node[shape=rectangle, draw=red, fill=red, minimum width=1.5cm] (fifthbox) [right=of fourthbox] {$e_{b}^{2}$};  
\node[shape=rectangle, draw=red, fill=red, minimum width=1.5cm] (sixthbox) [right=of fifthbox] {$e_{b}^{3}$};  
\draw[->] (firstbox.south) .. controls +(down:7mm) and +(left:.5mm).. (fourthbox.south);    
\draw[->] (secondbox.south) .. controls +(down:9mm)and +(left:.5mm).. (fifthbox.south);  
\draw[->] (thirdbox.south) .. controls +(down:11mm) and +(left:.5mm)..(sixthbox.south);  
\end{tikzpicture}

\medskip{}

Note that for any location vector, if the capacity transfer scheme
is a vector of zeros, the SSPwCT choice rules are equivalent to the
slot-specific priorities choice rules in \emph{Kominers and Sönmez}
(2016).

\subsection*{Substitutable Completion of SSPwCT Choice Rules }

A choice function $C^{b}$ satisfies the \textbf{irrelevance of rejected
contracts} (IRC) condition if for all $Y\subset X$, for all $z\in X\setminus Y$,
and $z\notin C^{b}\left(Y\cup\left\{ z\right\} \right)$ implies $C^{b}\left(Y\right)=C^{b}\left(Y\cup\{z\}\right)$.
A choice function $C^{b}$ satisfies \textbf{substitutability }if
for all $z,\,z'\in X$, and $Y\subseteq X$, $z\notin C^{b}\left(Y\cup\left\{ z\right\} \right)\Longrightarrow z\notin C^{b}\left(Y\cup\left\{ z,z'\right\} \right)$.
A choice function $C^{b}$ satisfies the \textbf{law of aggregate
demand }(LAD) if $Y\subseteq Y'\Longrightarrow\mid C^{b}\left(Y\right)\mid\,\leq\,\mid C^{b}\left(Y'\right)\mid$.

The following definitions are from \emph{Hatfield and Kominers }(2019): 
\begin{quotation}
A \emph{$\mathbf{completion}$ }of a choice function $C^{b}$ of branch
$b\in\mathcal{B}$ is a choice function $\overline{C}^{b}$, such
that for all $X\subseteq\mathcal{X}$, either $\overline{C}^{b}(X)=C^{b}(X)$
or there exists a distinct $x,x^{'}\in\overline{C}^{b}(X)$ such that
$\mathbf{i}(x)=\mathbf{i}(x^{'})$. If a choice function $C^{b}$
has a completion that satisfies the substitutability and \emph{IRC}
condition, then $C^{b}$ is said to be\emph{ $\mathbf{substitutably}$
$\mathbf{completable}$. }If every choice function in a profile $C=(C^{b})_{b\in\mathcal{B}}$
is substitutably completable, then we say that $C$ is \emph{$\mathbf{substitutably}$
$\mathbf{completable}$.}
\end{quotation}
Given the precedence order of slots $\triangleright_{b}$, the capacity
transfer function $q_{b}$, and the priority profile of slots $\Pi_{b}$,
and a set of contracts $X\subseteq\mathcal{X}$, we define a related
choice procedure $\overline{C}^{b}$. To formulate this related choice
rule, we first rename the slots as $S=(s^{1},s^{2},...,s^{2n_{b}}$)
where $s^{k}$ is either an original or a shadow seat, depending on
$\triangleright_{b}^{O}$ , $\triangleright_{\text{b}}^{E}$ , and
$L_{b}$. 
\begin{itemize}
\item Start with the original seat $s_{b}^{1}$. Assign the contract $x^{1}$
that is $\Pi_{b}^{s_{b}^{1}}-maximal$ among the contracts in $X$. 
\item If $s_{b}^{2}$ is either an original or a shadow seat such that $\mathbf{1}_{o_{b}^{1}}=0$
and $q_{b}^{1}=1$, assign the contract $x^{2}$ that is $\Pi_{b}^{s_{b}^{2}}-maximal$
among the contracts in $X\setminus\{x^{1}\}$. Otherwise, assign the
empty set.
\item This process continues in sequence. If $s_{b}^{k}$ is an original
seat or a shadow seat such that $\mathbf{1}_{o_{b}^{r}}=0$, where
$o_{b}^{r}$ is the original seat that is associated with the shadow
seat $s_{b}^{k}$, and $q_{b}^{r}=1$, then assign contract $x^{k}$
that is $\Pi_{b}^{s_{b}^{k}}-maximal$ among the contracts in $X\setminus\{x^{1},...,x^{k-1}\}$.
Otherwise, assign the empty set.
\end{itemize}
The difference between the SSPwCT choice rule $C^{b}$ we defined
in the main text and $\overline{C}^{b}$ defined above is as follows:
In the computation of $C^{b}$, if a contract of an agent is chosen
by some slot, then her other remaining contracts are removed for the
rest of the choice procedure. On the other hand, in the computation
of $\overline{C}^{b}$, if a contract of an agent is chosen by a slot,
then her other contracts will still be available for the following
slots. 

The following proposition shows that $\overline{C}^{b}$ defined above
is the completion of the SSPwCT choice rule $C^{b}$. 
\begin{prop}
$\overline{C}^{b}$ is a completion of $C^{b}$. 
\end{prop}
Our next results shows that $\overline{C}^{b}$ satisfies the IRC
condition, the substitutability and the LAD. 
\begin{prop}
$\overline{C}^{b}$ is substitutable, satisfies the IRC condition,
and the LAD. 
\end{prop}

\subsection{Cumulative Offer Mechanism }

A mechanism $\Psi(\cdot,C)$, where $C=(C^{b})_{b\in\mathcal{B}}$
is a given profile of choice rules for branches, is a mapping from
preference profiles of agents $P=(P_{i})_{i\in I}$ to outcomes. A
mechanism $\Psi(\cdot,C)$ is \emph{stable} if $\Psi(P,C)$ is a stable
outcome for every preference profile $P$. A mechanism $\Psi(\cdot,C)$
is \emph{strategy-proof} if for every preference profile $P$, and
for each individual $i\in\mathcal{I}$, there is no reported preference
$\widetilde{P}_{i}$, such that 
\[
\Psi((\widetilde{P}_{i},P_{-i}),C)P_{i}\Psi(P,C).
\]

We now introduce the cumulative offer mechanism (COM), whose outcome
is found with the following cumulative offer algorithm: 

\paragraph*{Step 1. }

An arbitrarily chosen agent propose her first choice contract $x_{1}$.
The branch $\mathbf{b}(x_{1})$ holds the contract $x_{1}$ if $C^{\mathbf{b}(x_{1})}\left(\left\{ x_{1}\right\} \right)=\left\{ x_{1}\right\} $,
and rejects it otherwise. Let $A_{\mathbf{b}(x_{1})}^{1}=\left\{ x_{1}\right\} $,
and $A_{b}^{1}=\emptyset$ for all $b\in\mathcal{B}\setminus\left\{ \mathbf{b}(x_{1})\right\} $. 

In general,

\paragraph*{Step t.}

An arbitrarily chosen agent, for whom no branch currently holds a
contract, proposes her favorite contract, call it $x_{t}$, among
the ones that have not been rejected in the previous steps. The branch
$\mathbf{b}(x_{t})$ holds $x_{t}$ if $x_{t}\in C^{\mathbf{b}(x_{t})}\left(A_{\mathbf{b}(x_{t})}^{t-1}\cup\left\{ x_{t}\right\} \right)$
and rejects it otherwise. Let $A_{\mathbf{b}(x_{t})}^{t}=A_{\mathbf{b}(x_{t})}^{t-1}\cup\left\{ x_{t}\right\} $,
and $A_{b}^{t}=A_{b}^{t-1}$ for all $b\in\mathcal{B}\setminus\left\{ \mathbf{b}(x_{t})\right\} $.

This algorithm terminates when every agent is matched to a branch
or every unmatched agent has all acceptable contracts rejected. Since
there are finitely many contracts, the algorithm terminates in some
finite step $T$. The final outcome is $\underset{b\in\mathcal{B}}{\cup}C^{b}\left(A_{b}^{T}\right)$. 
\begin{thm}
The COM with respect to SSPwCT choice rules  is stable and strategy-proof. 
\end{thm}

\section{Respecting Improvements }

Respect for improvements is an intuitive and desirable property in
practice. It suggest that agents should have no incentive to try to
lower their standings in branches' priority orders. This natural property
becomes crucial, especially in merit-based systems where branches'
priority orderings are determined through exam scores. To formally
define it in our framework, fix the precedence order $\triangleright_{b}\equiv(\triangleright_{b}^{O},\triangleright_{b}^{E},L_{b})$
and the capacity transfer function $q_{b}$ of branch $b$. 
\begin{defn}
We say that a choice rule $C^{b}(\cdot,\triangleright_{b},q_{b},\overline{\Pi}_{b})$
of branch $b$ is an\emph{ improvement over} $C^{b}(\cdot,\triangleright_{b},q_{b},\Pi_{b})$\emph{
}for agent $i\in\mathcal{I}$\emph{ }if for all slots $s\in O_{b}\cup E_{b}$
the following conditions hold:
\end{defn}
\begin{enumerate}
\item for all $x\in\mathcal{X}_{i}$ and $y\in\left(\mathcal{X}_{\mathcal{I}\setminus\{i\}}\cup\{\emptyset\}\right)$,
if $x\Pi_{b}^{s}y$ then $x\overline{\Pi}_{b}^{s}y$; and 
\item for all $y,z\in\mathcal{X}_{\mathcal{I}\setminus\{i\}}$, $y\Pi_{b}^{s}z$
if and only if $y\overline{\Pi}_{b}^{s}z$. 
\end{enumerate}
That is, $C^{b}(\cdot,\triangleright_{b},q_{b},\overline{\Pi}_{b})$
is an improvement over $C^{b}(\cdot,\triangleright_{b},q_{b},\Pi_{b})$
for agent $i$ if $\overline{\Pi}_{b}$ is obtained from $\Pi_{b}$
by increasing the priority of some of $i$'s contracts while leaving
the relative priority of other agents' contracts unchanged. We say
that a profile of branch choices $\overline{C}\equiv(C^{b}(\cdot,\triangleright_{b},q_{b},\overline{\Pi}_{b}))_{b\in\mathcal{B}}$
is an $\mathbf{improvement}$ over $C\equiv(C^{b}(\cdot,\triangleright_{b},q_{b},\Pi_{b}))_{b\in\mathcal{B}}$
for agent $i\in\mathcal{I}$ if, for each branch $b\in\mathcal{B}$,
$C^{b}(\cdot,\triangleright_{b},q_{b},\overline{\Pi}_{b})$ is an\emph{
}improvement over\emph{ }the choice rule $C^{b}(\cdot,\triangleright_{b},q_{b},\Pi_{b})$. 

We say that a mechanism $\varphi$ \textbf{$\mathbf{respects}$ $\mathbf{improvements}$}
for $i\in I$ if for any preference profile $P\in\times_{i\in I}\mathcal{P}^{i}$
\[
\varphi_{i}(P;\overline{C})R^{i}\varphi_{i}(P;C)
\]
 whenever $\overline{C}\equiv(C^{b}(\cdot,\triangleright_{b},q_{b},\overline{\Pi}_{b}))_{b\in\mathcal{B}}$
is an improvement over $C\equiv(C^{b}(\cdot,\triangleright_{b},q_{b},\Pi_{b}))_{b\in\mathcal{B}}$
. 
\begin{thm}
The COM with respect to SSPwCT choice rules respects improvements. 
\end{thm}

\section{Comparative Statics }

In this section, we first look at the effect of increasing the transferability
of original seats on agents' welfare under the COM with respect to
SSPwCT choice rules. We, then, extend the comparative static results
of \emph{Kominers }(2020) to the SSPwCT family. \emph{Kominers }(2020)
provides a new proof of the entry comparative static, by way of the
respect for improvements property. The author sheds light on a strong
relationship between several different entry comparative statics and
the respecting improvement property in many-to-one matching markets
with contracts. Building on his formulation, we analyze the effect
of expanding the capacity of a single branch on agents' welfare agent-proposing
under the COM in SSPwCT environment. We also examine the effect of
adding contracts on agents' welfare under the COM in our setting. 

\subsection{Increasing Transferability}

SSPwCT is a large family of choice rules. If transferability of all
original slots is prohibited, we obtain the slot-specific priorities
choice rules of \emph{Kominers and Sönmez }(2016). Allowing transferability
of an original slot, while everything else remains fixed, is welfare-improving
for agents. We analyze the monotonicity of improvements on agents'
welfare by only changing the transferability of original slots, while
all else remains fixed. 

Let $\widetilde{q}_{b}$ and $q_{b}$ be two capacity transfer schemes
for branch $b$. We say that $\widetilde{q}_{b}$ is $\mathbf{more}$
$\mathbf{flexible}$ than $q_{b}$ if $\widetilde{q}_{b}>q_{b}$,
i.e., $\widetilde{q}_{b}^{k}\geq q_{b}^{k}$ for all $k=1,...,n_{b}$
and $\widetilde{q}_{b}^{l}>q_{b}^{l}$ for some $l=1,...,n_{b}$.
Suppose that $\triangleright_{b}$ and $\Pi_{b}$ are fixed. Then,
the SSPwCT choice rule $C^{b}(\cdot,\triangleright_{b},\widetilde{q}_{b},\Pi_{b})$
can be interpreted as an expansion of the SSPwCT choice rule $C^{b}(\cdot,\triangleright_{b},q_{b},\Pi_{b})$
if $\widetilde{q}_{b}$ is more flexible than $q_{b}$. We are now
ready to present our result. 
\begin{thm}
Suppose that $Z$ is the outcome of the COM  at $(P,C)$, where $P=(P_{i_{1}},...,P_{i_{n}})$
is the profile of agent preferences and $C=(C^{b_{1}},...,C^{b_{m}})$
is the profile of branches' SSPwCT choice rules. Fix a branch $b\in\mathcal{B}$.
Suppose that $\widetilde{C}^{b}$ takes as an input capacity transfer
scheme that is more flexible than that of $C^{b}$, holding all else
constant. Then, the outcome of the COM  at $(P,(\widetilde{C}^{b},C_{-b}))$,
$\widetilde{Z}$, Pareto dominates $Z$. 
\end{thm}
Theorem 3 is intuitive and indicates that making an untransferable
original slot of a branch transferable leads to strategy-proof Pareto
improvement of the COM. One should note that expanding a branch's
choice rule changes the stability definition. However, Theorem 3 provides
a normative foundation for such a change, as it increases agents'
welfare.\footnote{This result does not contradict the findings of \emph{Alva and Manjunath
}(2019), because transferring the capacity of an original slot to
an associated shadow slot changes the feasible set in their context.} 

\subsection{Expanding Capacity}

We follow the formulation of \emph{Kominers }(2020) in this section.
We look at how the COM outcome changes when an original slot is added
to branch $b$ in the SSPwCT environment. Suppose the choice rules
of all branches other than $b$ are fixed. We extend the set of original
slots in branch $b$ from $O_{b}$ to $\widetilde{O}_{b}=O_{b}\cup\{\widetilde{o}\}$,
where $\widetilde{o}$ is the newly added original slot. We assume
there is no change in the priorities of slots in $O_{b}$. We write
$\widetilde{\Pi}_{b}=(\Pi_{b},\pi^{\widetilde{o}})$, where $\Pi_{b}=(\Pi_{b}^{o},\Pi_{b}^{e})$
is the priority profile of slots in $O_{b}\cup E_{b}$ and $\pi^{\widetilde{o}}$
is the priority ordering of new original slot $\widetilde{o}$. 

As pointed out by \emph{Kominers }(2020), adding a new original slot
$\widetilde{o}$ is similar to boosting the ranking of contracts of
all agents that were deemed unacceptable in $\pi^{\widetilde{o}}$,
keeping constant all other slots' rankings. Hence, by our Theorem
2, adding $\widetilde{o}$ leads to an improvement for all agents.
We state this result formally as follows: 
\begin{thm}
Let $Z$ and $\widetilde{Z}$ be the outcomes of the COM in the markets
with the set of slots $\{O_{b}\cup E_{b}\}_{b\in\mathcal{B}}$ and
$\{O_{b}\cup E_{b}\}_{b\in\mathcal{B}}\cup\{\widetilde{o}_{b}\}$,
respectively, where $\widetilde{o}_{b}$ is an original slot added
to branch $b$. Then, each agent $i\in\mathcal{I}$ (weakly) prefers
her assignment under $\widetilde{Z}$ to her assignment under $Z$. 
\end{thm}
This result expands Theorem 2 of \emph{Kominers }(2020) to the SSPwCT
family. Our Theorems 3 and 4 together imply the following: if a new
original slot and its associated shadow slot are added and transferability
is allowed, then the COM outcome is improved for agents. 

\subsection{Adding Contracts}

\emph{Kominers (2020)} also shows that adding new contracts at the
bottom of some slots' priority orders (that is, right before the null
contract $\emptyset$) improves outcomes for all agents. We follow
his terminology to extend his result to the SSPwCT family. The following
formulation is adapted to the SSPwCT environment from \emph{Kominers
}(2020): Let $X$ be an initial set of contracts and $\widetilde{X}$
is a newly added set of contracts, expanding the set to $X\cup\widetilde{X}$.
Let $\widetilde{P}=(\widetilde{P}_{i})_{i\in\mathcal{I}}$ and $\widetilde{\Pi}=(\widetilde{\Pi}_{b})_{b\in\mathcal{B}}$
denote the vector of agent preferences and slot priorities over $X\cup\widetilde{X}$,
respectively. Suppose that (1) $x\widetilde{P}_{i}x^{'}$ if and only
if $xP_{i}x^{'}$ for all $i\in\mathcal{I}$ and (2) $x,x^{'}\in X$,
and $x\widetilde{\Pi}_{b}^{s}x^{'}$ if and only if $x\Pi_{b}^{s}x^{'}$
for all slots $s\in\{O_{b}\cup E_{b}\}_{b\in\mathcal{B}}$ and $x,x^{'}\in X$.
If $x\widetilde{\Pi}_{b}^{s}\widetilde{x}$ for all slots $s\in\{O_{b}\cup E_{b}\}_{b\in\mathcal{B}}$,
$x\in X$, and $\widetilde{x}\in\widetilde{X}$, then $\widetilde{\Pi}$
is an \emph{improvement} over $\Pi$ under $X\cup\widetilde{X}$.
Each slot $s\in\{O_{b}\cup E_{b}\}_{b\in\mathcal{B}}$ deems all the
contracts in $\widetilde{X}$ as unacceptable in $\Pi_{b}^{s}$. Hence,
$\widetilde{\Pi}$ can be obtained from $\Pi$ by improving the ranking
of contracts in $\widetilde{X}$ above the outside option. Then, our
Theorem 2 implies the following result. 
\begin{thm}
Let $Z$ and $\widetilde{Z}$ be the outcomes of the COM  in the markets
with the set of contracts $X$ and $X\cup\widetilde{X}$, respectively.
Then, each agent $i\in\mathcal{I}$ (weakly) prefers her assignment
under $\widetilde{Z}$ to her assignment under $Z$. 
\end{thm}
Our Theorem 5 expands Theorem 3 of \emph{Kominers }(2020) to the SSPwCT
family. 

In our final comparative static, we show that an agent receives a
(weakly) better outcome in the COM when new contracts are added for
this agent. The following formulation is adapted to the SSPwCT environment
from \emph{Kominers }(2020): Let $\widetilde{x}$ be a new contract
that is added to the contract set $X$, expanding it to $X\cup\{\widetilde{x}\}$.
Let $\widetilde{P}=(\widetilde{P}_{i})_{i\in\mathcal{I}}$ and $\widetilde{\Pi}=(\widetilde{\Pi}_{b})_{b\in\mathcal{B}}$
denote the vector of agent preferences and slot priorities over $X\cup\{\widetilde{x}\}$,
respectively. Suppose (1) $x\widetilde{P}_{i}x^{'}$ if and only if
$xP_{i}x^{'}$ for all $i\in\mathcal{I}$ and $x,x^{'}\in X$, and
(2) $x\widetilde{\Pi}_{b}^{s}x^{'}$ if and only if $x\Pi_{b}^{s}x^{'}$
for all slots $s\in\{O_{b}\cup E_{b}\}_{b\in\mathcal{B}}$ and $x,x^{'}\in X$.
Then, our Theorem 2 implies the following result. 
\begin{thm}
Let $Z$ and $\widetilde{Z}$ be the outcomes of the COM  in the market
with the set of contracts $X$ and $X\cup\{\widetilde{x}\}$, respectively.
Then, each agent $\mathbf{i}(\widetilde{x})$ (weakly) prefers her
assignment under $\widetilde{Z}$ to her assignment under $Z$. 
\end{thm}
This result expands Theorem 4 of \emph{Kominers }(2020) to the SSPwCT
family. 

\section{Conclusions}

This paper introduces a practical family of SSPwCT choice rules. We
show that when these choice rules are used by institutions the COM
is stable, strategy-proof, and respects improvements.  Moreover, we
show that transferring the capacity of one more unfilled slot, if
all else remains constant, leads to strategy-proof Pareto improvement
of the COM. We also show that both expansion of branch capacities
and adding new contracts (weakly) increase agents' welfare under the
COM. 

The SSPwCT choice rules expands the toolkit available to market designers
and may be used in real-world matching markets to accommodate diversity
concerns. We believe SSPwCT choice rules may be invoked in cadet-branch
matching in USMA and ROTC, resource allocation problems in India with
comprehensive affirmative action, Chilean school choice with affirmative
action constraints, and Brazilian college admissions with multidimensional
reserves, among many others.

\section*{Appendices }

\paragraph*{Proof of Proposition 1. }

Let $\triangleright_{b}$ the precedence order of slots, $q_{b}$
be the capacity transfer function, and $\Pi_{b}$ be the priority
profile of slots. Consider an offer set $X\subseteq\mathcal{X}$.
If there are distinct $x,x^{'}\in\overline{C}^{b}\left(X\right)$
such that $\mathbf{i}(x)=\mathbf{i}(x^{'})$, then we are done. Now,
suppose that there is no pair of contracts $x,x^{'}\in\overline{C}^{b}\left(X\right)$
such that $\mathbf{i}(x)=\mathbf{i}(x^{'})$. We need to show that
$C^{b}\left(X\right)=\overline{C}^{b}\left(X\right)$ in this case. 

We prove our claim by induction on slots' indexes $k=1,...,2n_{b}$.
We show that for each slot the contracts chosen by $C^{b}$ and $\overline{C}^{b}$
is the same. For the first slot, $s_{b}^{1}$, both $C^{b}$ and $\overline{C}^{b}$
chooses the $\Pi_{b}^{s_{b}^{1}}-maximal$ among the contracts in
$X$. By way of induction, assume that each of the first $k-1$ slots,
$s_{b}^{1},...,s_{b}^{k-1}$, selects the same contract under choice
procedures $C^{b}$ and $\overline{C}^{b}$, respectively. Call these
contracts $x^{1},...,x^{k-1}$, with the possibility that some of
$x^{j}$s, $j=1,....,k-1$, might be the null contract $\emptyset$.
We need to show that the $k^{th}$ slot $s_{b}^{k}$, selects the
same contract under $C^{b}$ and $\overline{C}^{b}$, respectively.
Let $x^{k}$ be the contract $C^{b}$ selects at slot $k$ among the
contracts in $X\setminus X_{\mathbf{i}(\{x^{1},...,x^{k-1}\})}$.
That is, $x^{k}$ is the $\Pi_{b}^{s_{b}^{k}}-maximal$ contract among
contracts in $X\setminus X_{\mathbf{i}(\{x^{1},...,x^{k-1}\})}$.
Under $\overline{C}^{b}$, slot $s_{b}^{k}$ selects $\Pi_{b}^{s_{b}^{k}}-maximal$
contract among $X\setminus\{x^{1},...,x^{k-1}\}$. Since $\Pi_{b}^{s_{b}^{k}}$
is a linear order and no contract of agents $\mathbf{i}\left(\left\{ x^{1},...,x^{k-1}\right\} \right)$
is chosen at any other slot by our initial supposition, $\Pi_{b}^{s_{b}^{k}}-maximal$
contract among the contracts in $X\setminus\{x^{1},...,x^{k-1}\}$
is $x^{k}$. Hence, the same contract $x^{k}$ is chosen at slot $s_{b}^{k}$
under both $C^{b}$ and $\overline{C}^{b}$. This completes the proof
of Proposition 1. 

\paragraph*{Proof of Proposition 2. }

We first show that $\overline{C}^{b}$ satisfies the IRC condition.
For any $X\subseteq\mathcal{X}$ such that $X\neq\overline{C}^{b}(X)$,
let $x$ be one of the rejected contracts, i.e., $x\in X\setminus\overline{C}^{b}(X)$.
We need to show that 
\[
\overline{C}^{b}(X)=\overline{C}^{b}(X\setminus\{x\}).
\]
 Let $x^{k}$ and $\widetilde{x}^{k}$ be the contracts chosen by
the slot $s_{b}^{k}$ under $\overline{C}^{b}$ from the set $X$
and $X\setminus\{x\}$, respectively. We will show that $x^{k}=\widetilde{x}^{k}$
for each $k=1,...,2n_{b}$ by induction. For the first slot $s_{b}^{1}$,
since we know that $x$is not chosen by $\overline{C}^{b}$---and,
hence by $s_{b}^{1}$--- and $\Pi^{s_{b}^{1}}$ is a linear order
we have $x^{1}=\widetilde{x}^{1}$. Suppose that $x^{j}=\widetilde{x}^{j}$
for $j=1,...,k-1$. We need to show that $x^{k}=\widetilde{x}^{k}$.
Notice that the set of remaining contracts for slot $s_{b}^{k}$ from
the choice processes starting with $X$ and $X\setminus\{x\}$ are
$X\setminus\{x^{1},...,x^{k-1}\}$ and $X\setminus\{x,x^{1},...,x^{k-1}\}$,
respectively. Since we know that $x$ is not chosen, and hence, is
not $\Pi_{b}^{s_{b}^{k}}-maximal$ among the contracts $X\setminus\{x^{1},...,x^{k-1}\}$,
then we have $x^{k}=\widetilde{x}^{k}$. Therefore, at each slot,
the same contract is chosen from the choice processes starting with
$X$ and $X\setminus\{x\}$, respectively. This completes our proof. 

\paragraph*{Substitutability. }

Consider an offer set $X\subseteq\mathcal{X}$ such that $X\neq\overline{C}^{b}(X)$.
Let $x$ be one of the rejected contracts, i.e., $x\in X\setminus\overline{C}^{b}(X)$,
and let $z$ be an arbitrary contract in $\mathcal{X}\setminus X$.
To show substitutability, we need to show that 
\[
x\notin\overline{C}^{b}(X\cup\{z\}).
\]
There are two cases to consider: 

\paragraph*{\emph{Case 1:}}

$z\notin\overline{C}^{b}\left(X\cup\{z\}\right)$. 

Since $\overline{C}$ satisfies the IRC condition, we have 
\[
\overline{C}^{b}(X)=\overline{C}^{b}(X\cup\{z\}).
\]

Since $x\notin\overline{C}^{b}(X)$ is given, we have $x\notin\overline{C}^{b}(X\cup\{z\})$,
which is the desired conclusion. 

\paragraph*{\emph{Case 2: }}

$z\in\overline{C}^{b}\left(X\cup\{z\}\right)$. 

We call the choice processes starting with the contract sets $X$
and $X\cup\{z\}$, respectively, as the $\mathbf{initial}$ and $\mathbf{new}$
processes. Note that the precedence order $\triangleright_{b}$, the
capacity transfer scheme $q_{b}$, and the priority profile of slots
$\Pi_{b}$ are the same in the initial and new processes. Let $Y^{i}$
and $\overline{Y}^{i}$ be the sets of remaining contracts after slot
$s_{b}^{i}$, $i=1,...,2n_{b}-1$, selects its contract in the initial
and new processes, respectively.

Let $s_{b}^{j}$ be the slot that selects the contract $z$ in the
new process. For each slot $i=1,...,j-1$, we have $\overline{Y}^{i}=Y^{i}\cup\{z\}$
as each slot $s_{b}^{i}$, $i=1,...,j-1$, selects the same contract
under the initial and new processes. Since slot $s_{b}^{j}$ selects
$z$, $x$ is not chosen in any of $s_{b}^{1},...,s_{b}^{j}$ in the
new process. We have the following possibilities: 
\begin{enumerate}
\item $s_{b}^{j}$ is an \emph{original} seat with $q_{b}^{j}=0$ and selects
$\emptyset$ in the initial process. 
\item $s_{b}^{j}$ is an \emph{original} seat with $q_{b}^{j}=1$ and selects
$\emptyset$ in the initial process.
\item $s_{b}^{j}$ is an \emph{original} seat and selects a different contract---call
it $y$---in the initial process. 
\item $s_{b}^{j}$ is a\emph{ shadow} seat and selects $\emptyset$ in the
initial process.
\item $s_{b}^{j}$ is a\emph{ shadow} seat and selects a different contract---call
it $y$---in the initial process.
\end{enumerate}
In cases 1 and 4, $Y^{j}=\overline{Y}^{j}$ and the rest of the initial
and the new process will be the same. Therefore, for these cases we
can conclude that $x\notin\overline{C}\left(X\cup\{z\}\right)$. In
case 2, $Y^{j}=\overline{Y}^{j}$ and the associated shadow seat of
$s_{b}^{j}$ is passive in the new process while it is active in the
initial process. In cases 3 and 5, $Y^{j}\cup\{y\}=\overline{Y}^{j}$.
Note that in case 5, the associated shadow seat of $s_{b}^{j}$ is
passive as $\mathbf{1}_{s_{b}^{j}}=1$.

For every seat $s_{b}^{\kappa}$, $\kappa=j,...,2n_{b}$, we have
either $(i)$ $Y^{\kappa}=\overline{Y}^{\kappa}$ and the capacity
of $s_{b}^{k+1}$ is either the same in the initial and new processes
or 1 in the initial process and 0 in the new process, or $(ii)$ $Y^{\kappa}\cup\{y\}=\overline{Y}^{\kappa}$,
and the capacity of $s_{b}^{k+1}$ is the same under both processes.
We have already showed it for $\kappa=j$ above. By the way of induction,
suppose that the assertion holds for slots $\kappa=j,...,k-1$. We
need to show that it holds for $\kappa=k$. 

If $Y^{k-1}=\overline{Y}^{k-1}$ and the capacity of the $k^{th}$
slot is 0 in both initial and new processes, then we will have $Y^{k}=\overline{Y}^{k}$,
and the capacity of $s_{b}^{k+1}$ will be the same under the initial
and the new processes. Similarly, if $Y^{k-1}=\overline{Y}^{k-1}$
and the capacity of $s_{b}^{k}$ is 1 in both initial and new processes,
then the same contract will be chosen at $s_{b}^{k}$ . Hence, we
will have $Y^{k}=\overline{Y}^{k}$, and the capacity of $s_{b}^{k+1}$
will be the same under the both processes. 

Now suppose $Y^{k-1}=\overline{Y}^{k-1}$ and the capacity of $s_{b}^{k}$
is 1 in the initial process and 0 in the new processes. Note that
$s_{b}^{k}$ must be a shadow seat in this case. If seat $s_{b}^{k}$
selects $\emptyset$ in the initial process, then we have $Y^{k}=\overline{Y}^{k}$
and the capacity of $s_{b}^{k+1}$ is the same under both process.
If seat $s_{b}^{k}$ selects contract $y$ in the initial process,
then we have $Y^{k}\cup\{y\}=\overline{Y}^{k}$. Moreover, the capacity
of $s_{b}^{k+1}$ is the same under both processes. 

Finally, suppose $Y^{k-1}\cup\{y\}=\overline{Y}^{k-1}$, and the capacity
of $s_{b}^{k}$ is the same under the both processes. If the capacity
of $s_{b}^{k}$ is 0, then $Y^{k-1}\cup\{y\}=\overline{Y}^{k-1}$,
and the capacity of $s_{b}^{k+1}$ is the same under the both processes.
If the capacity of $s_{b}^{k}$ is 1 and $Y^{k-1}\cup\{y\}=\overline{Y}^{k-1}$,
either the same contract is chosen under both processes, or $\emptyset$
is chosen in the initial process and $y$ is chosen in the new process.
In the former case, we have $Y^{k}\cup\{z\}=\overline{Y}^{k}$ for
some contract $z$ and the capacity of $s_{b}^{k+1}$ is the same
under both processes. In the latter case, $Y^{k}=\overline{Y}^{k}$
and the capacity of $s_{b}^{k+1}$ is either the same under both processes---in
the cases, where $s_{b}^{k}$ is a shadow seat or is an original seat
with $q_{b}^{k}=0$---or the capacity of $s_{b}^{k+1}$ is 1 in the
initial process and 0 in the new process---in the case where $s_{b}^{k}$
is an original slot with $q_{b}^{k}=1$. This ends the proof of our
claim. 

Since $x$ is not chosen by any seat in the initial process, it will
not be chosen under the new process either as a result of the above
claim. Therefore, $\overline{C}^{b}$ is substitutable. 

\paragraph*{Law of Aggregate Demand. }

Consider two sets of contracts $X$ and $Y$ such that $X\subseteq Y\subseteq\mathcal{X}$.
We want to show that 
\[
\mid\overline{C}^{b}(X)\mid\leq\mid\overline{C}^{b}(Y)\mid.
\]
 We call the choice process starting with the contract set X as ``process
X'', and the choice process starting with the contract set Y as ``process
Y''. Let $X^{j}$ and $Y^{j}$ be the set of remaining contracts
after slot $s_{b}^{j}$ selects its contract in processes X and Y,
respectively, for $j=1,...,2n_{b}$. We will prove that for each $j=1,...,2n_{b}$,
$X^{j}\subseteq Y^{j}$. 

The first slot $s_{b}^{1}$ is an original slot. Let $x^{1}$ be the
contract chosen by $s_{b}^{1}$ in process X. Since $X\subset Y$,
$x^{1}\in Y$. In process Y, $s_{b}^{1}$ either selects $x^{1}$
or another contract $y\in Y\setminus X$. In both cases, we have $X^{1}\subseteq Y^{1}$.
By the way of induction, suppose that $X^{j}\subseteq Y^{j}$, for
all $j=1,...,k-1$. We need to show $X^{k}\subseteq Y^{k}$. 

First suppose that $s_{b}^{k}$ is an original seat. Since $X^{k-1}\subseteq Y^{k-1}$,
if $s_{b}^{k}$ chooses a contract $x^{k}\in X^{k-1}\subseteq Y^{k-1}$
in process X, then $s_{b}^{k}$ chooses either $x^{k}$ or another
contract $z\in Y^{k-1}\setminus X^{k-1}$. In both cases, $X^{k}\subseteq Y^{k}$. 

Now suppose that, $s_{b}^{k}$ is a shadow seat. Note that the capacity
of the shadow seat in process X is either the same as the capacity
of this seat in process Y (both 0 or 1), or the capacity of $s_{b}^{k}$
is 1 in process X and 0 in process Y---that is, the associated original
seat is filled in process Y, but remained vacant in process X. If
they are both 0, then $X^{k-1}\subseteq Y^{k-1}$, and $X^{k-1}=X^{k}$
and $Y^{k-1}=Y^{k}$ imply $X^{k}\subseteq Y^{k}$. If they are both
1 and the contract $x^{k}$ is chosen by seat $s_{b}^{k}$ in process
X, then either contract $x^{k}$ or another contract $w\in Y^{k-1}\setminus X^{k-1}$
is chosen in process Y. Both imply $X^{k}\subseteq Y^{k}$. Finally,
suppose that $s_{b}^{k}$ is active in process X and passive in process
Y. This is the case only when the associated original seat is filled
in process Y with a contract in $Y\setminus X$ and remains vacant
in process X. Given that $X^{k-1}\subseteq Y^{k-1}$, if a contract
$x^{k}$ is chosen in process X, we have $X^{k}\subseteq Y^{k}$.
It is important to note that it cannot be the case where the shadow
seat $s_{b}^{k}$ is passive in process X and active in process Y.
It contradicts with the inductive assumption. 

Since for all seats $s_{b}^{j}$, $j=1,...,2n_{b}$, we have $X^{j-1}\subseteq Y^{j-1}$,
then we can conclude that $\overline{C}^{b}(X)\mid\leq\mid\overline{C}^{b}(Y)\mid$. 

\paragraph{Proof of Theorem 2.}

Assume, toward a contradiction, that the COM with regard to SSPwCT
does not respect improvements. Then, there exists an agent $i\in I$,
a preference profile of agents $P\in\times_{i\in I}\mathcal{P}^{i}$,
and choice profiles $\overline{C}$ and $C$ such that $\overline{C}$
is an improvement over $C$ for agent $i$ and 
\[
\mathcal{C}_{i}(P;C)P^{i}\mathcal{C}_{i}(P;\overline{C}).
\]
 Let $\mathcal{C}_{i}(P;C)=x$ and $\mathcal{C}_{i}(P;\overline{C})=\overline{x}$.
Consider a preference $\widetilde{P}^{i}$ of agent $i$ according
to which the only acceptable contract is $x$, i.e., $\widetilde{P}^{i}:\:x-\emptyset_{i}$.
Let $\widetilde{P}=(\widetilde{P}^{i},P_{-i})$. We will first prove
the following claim:

\paragraph*{Claim: $\mathcal{C}_{i}(\widetilde{P};C)=x$ $\protect\Longrightarrow$
$\mathcal{C}_{i}(\widetilde{P};\overline{C})=x$. }

\paragraph*{Proof of the Claim: }

Consider the outcome of the COM under choice profile $C$ given the
preference profile of agents $\widetilde{P}$. By Hirata and Kasuya
(2014), the cumulative offer is order-independent. We can first completely
ignore agent $i$ and run the COM until it stops. Let $Y$ be the
outcome. At this point, agent $i$ makes an offer for his only contract
$x$. This might create a chain of rejections, but it does not reach
agent $i$ since we assumed $\mathcal{C}_{i}(\widetilde{P};C)=x$.
Let the $k^{th}$ slot with respect to precedence order $\triangleright_{\mathbf{b}(x)}$
be the slot that chose contract $x$. 

Now consider the COM  under choice profile $\overline{C}$. Again,
we completely ignore agent $i$ and run the COM until it stops. The
same outcome $Y$ is obtained, because the only difference between
the two COMs  is agent $i$'s position in the priority rankings. At
this point, agent $i$ makes an offer for his only contract $x$.
If $x$ is chosen by the same slot, i.e., $k^{th}$ slot with respect
to $\triangleright_{\mathbf{b}(x)}$, then the same rejection chain
(if there was one in the COM  under the choice profile $C$) will
occur and it does not reach agent $i$; otherwise, we would have a
contradiction with the case under choice profile $C$. The only other
possibility is the following: since agent $i$'s ranking is now (weakly)
better under $\overline{\Pi}_{\mathbf{b}(x)}$ compared to $\Pi_{\mathbf{b}(x)}$,
his contract $x$ might be chosen by slot $l$ which precedes slot
$k$ with respect to $\triangleright_{\mathbf{b}(x)}$. Then, it must
be the case that by selecting $x$ slot $l$ must reject some other
contract it was holding. Let us call this contract $y$. If no contract
of agent  $\mathbf{i}(y)=j$ is chosen between slots $l$ and $k$,
then the slots between $l$ and $k$ choose the same contracts under
both priority profiles. In this case, $y$ is chosen by slot $k$.
Thus, if a rejection chain starts, it will not reach agent $i$; otherwise,
we could have a contradiction, due to the fact that $x$ was chosen
at the end of the COM  under choice profile $C$. A different contract
of agent $j$ cannot be chosen between groups $l$ and $k$; otherwise,
the observable substitutability\footnote{By Hatfield, Kominers and Westkamp (2019), we know that any choice
function that has a substitutable and size monotonic completion must
be observably substitutable. Observable substitutability simply says
that branch choices along the cumulative offer process satisfy substitutes
property, i.e., for the offer sets that can arise during the cumulative
offer process. } of branch $\mathbf{b}(x)$'s SSPwCT choice rule would be violated.
Therefore, if any contract of agent $j$ is chosen by slots between
$l$ and $k$, it must be $y$. If $y$ is chosen by a slot that precedes
$k$, then it must replace a contract---we call this contract $z$.
By the same reasoning, no other contract of agent  $\mathbf{i}(z)$
can be chosen before slot $k$; otherwise, we would violate the observable
substitutability of branch $\mathbf{b}(x)$'s SSPwCT choice rule.
Proceeding in this fashion causes the same contract in slot $k$ to
be rejected and initiates the same rejection chain that occurs under
choice profile $C$. Since the same rejection chain does not reach
agent $i$ under choice profile $C$, it will not reach agent  $i$
under choice profile $\overline{C}$, which ends our proof for the
claim. 

Since $\mathcal{C}_{i}(P;C)=x$ and $\mathcal{C}_{i}(P;\overline{C})=\overline{x}$
such that $xP^{i}\overline{x}$, if agent $i$ misreports and submits
$\widetilde{P}^{i}$ under choice profile $\overline{C}$ , then she
can successfully manipulate the COM. This is a contradiction as we
have already established that the COM is strategy-proof. 

\paragraph{Proof of Theorem 3.}

Suppose that $Z$ is the outcome of the COM at $(P,C)$, where $P=(P_{i_{1}},...,P_{i_{n}})$
is the profile of agent preferences and $C=(C^{b_{1}},...,C^{b_{m}})$
is the profile of branches' SSPwCT choice rules. Consider a branch
$b\in\mathcal{B}$. Suppose that $\widetilde{C}^{b}$ and $C^{b}$
take as an input capacity transfer schemes $\widetilde{q}_{b}$ and
$q_{b}$, respectively, where $\widetilde{q}_{b}^{k}=q_{b}^{k}$ for
all $k=1,...,s-1,s+1,...,n_{b}$. For slot $s$, let $\widetilde{q}_{b}^{s}=1$
and $q_{b}^{s}=0$. That is, the capacity of the original seat $s$
is transferred to the associated shadow seat under capacity function
$\widetilde{q}_{b}$, but not under the capacity function $q_{b}$.
We need to prove that the outcome of the COM at $(P,(\widetilde{C}^{b},C_{-b}))$,
$\widetilde{Z}$, Pareto dominates $Z$. 

In the computation of COM, if the original slot $s$ is filled, then
we have $\widetilde{Z}=Z$ because, under both $\widetilde{q}_{b}$
and $q_{b}$ the shadow slot associated with the original slot $s$
will become inactive. 

We now consider the case where the original slot $s$ remains vacant
in the computation of COM under $(P,C)$. Then, under $(P,(\widetilde{C}^{b},C_{-b}))$,
the shadow slot associated with the original slot $s$ -- we call
it $\widetilde{s}$ -- will be active, i.e., it will have a capacity
of 1. There are two cases to consider. If the shadow slot $\widetilde{s}$
remains vacant in the computation of COM under $(P,(\widetilde{C}^{b},C_{-b}))$,
then we again have $\widetilde{Z}=Z$ , as the only difference between
the two COMs, under $(P,C)$ and $(P,(\widetilde{C}^{b},C_{-b}))$,
is the capacity of the shadow slot $\widetilde{s}$. 

The non-trivial case is the one where the shadow slot $\widetilde{s}$
is assigned a contract in COM under $(P,(\widetilde{C}^{b},C_{-b}))$.
We now define an \emph{improvements chain }algorithm that starts with
the outcome $Z$. 

\paragraph*{Step 1. }

Let $x_{1}$ be the contract that is assigned to slot $\widetilde{s}$
in the SSPwCT choice procedure of branch $\mathbf{b}(x_{1})$. If
agent $\mathbf{i}(x_{1})$ is assigned $\emptyset$ under $Z$, then
the improvement process ends and we have $\widetilde{Z}=Z\cup\{x_{1}\}$.
Otherwise, set as $z_{1}$ the contract that agent $\mathbf{i}(x_{1})$
is assigned under $Z$. Note that $x_{1}P_{\mathbf{i}(x_{1})}z_{1}$. 

\paragraph*{Step 2. }

Let $x_{2}$ be the contract that is chosen by the slot vacated by
$z_{1}$ (or the shadow seat that is associated with it). If agent
$\mathbf{i}(x_{2})$ is assigned $\emptyset$ under $Z$, then the
improvement process ends and we have $\widetilde{Z}=Z\cup\{x_{1},x_{2}\}\setminus\{z_{1}\}$.
Otherwise, set as $z_{2}$ the contract that agent $\mathbf{i}(x_{2})$
is assigned under $Z$. Note that $x_{2}P_{\mathbf{i}(x_{2})}z_{2}$.

\paragraph*{Step n. }

Let $x_{n}$ be the contract that is chosen by the slot vacated by
$z_{n-1}$ (or the shadow seat that is associated with it). If agent
$\mathbf{i}(x_{n})$ is assigned $\emptyset$ under $Z$, then the
improvement process ends and we have $\widetilde{Z}=Z\cup\{x_{1},...,x_{n}\}\setminus\{z_{1},...,z_{n-1}\}$.
Otherwise, set as $z_{n}$ the contract that agent $\mathbf{i}(x_{n})$
is assigned under $Z$. Note that $x_{n}P_{\mathbf{i}(x_{n})}z_{n}$. 

In every step of the improvement chain algorithm a contract is replaced
by a more preferred contract. Since there are finitely many contracts
the improvement chain algorithm must end. Therefore, we reach $\widetilde{Z}$,
which Pareto dominates $Z$, in finitely many step. This ends our
proof. 
\end{document}